\documentclass[namedreferences]{solarphysics}

\usepackage[hyperref,optionalrh]{spr-sola-addons} % For Solar Physics 
\usepackage{graphicx}        % For eps figures, newer & more powerfull
\usepackage{color}           % For color text: \color command
\usepackage{breakurl}        % For breaking URLs easily trough lines
\usepackage{xfrac}
\chardef\us=`\_

\begin{document}

\begin{article}
\begin{opening}

\title{The SWAP Filter: A Simple Azimuthally Varying Radial Filter for Wide-Field EUV Solar Images}

%\author[addressref={aff1},corref,email={e-mail.a@mail.com}]{\inits{F.N.}\fnm{Daniel B.}~\lnm{Seaton}}%\sep
%\author[addressref=aff1,email={e-mail.b@mail.com}]{\inits{F.}\fnm{Fisrt}~\lnm{Author-b}}%\sep
%\author[corref,email={e-mail.c@mail.com}]{\inits{S.}\fnm{Second}~\lnm{Author-c}}%\sep
%\author{\inits{T.}\fnm{Third}~\lnm{Author-x}}
\author[addressref={SwRI},corref,email={dseaton@boulder.swri.edu}]{\inits{D.B.}\fnm{Daniel~B.}~\lnm{Seaton}\orcid{0000-0002-0494-2025}}
\author[addressref={ROB}]{\inits{D.}\fnm{David}~\lnm{Berghmans}\orcid{0000-0003-4052-9462}}
\author[addressref={NU}]{\inits{D.S.}\fnm{D.~Shaun}~\lnm{Bloomfield}\orcid{0000-0002-4183-9895}}
\author[addressref={ESAC}]{\inits{A.}\fnm{Anik}~\lnm{De Groof}\orcid{0000-0002-0331-1029}}
\author[addressref={ROB}]{\inits{E.}\fnm{Elke}~\lnm{D'Huys}\orcid{0000-0002-2914-2040}}
\author[addressref={ROB}]{\inits{B.}\fnm{Bogdan}~\lnm{Nicula}\orcid{0000-0002-8050-5476}}
\author[addressref={NCEI}]{\inits{L.A.}\fnm{Laurel~A.}~\lnm{Rachmeler}\orcid{0000-0002-3770-009X}}
\author[addressref={SwRI}]{\inits{M.J.}\fnm{Matthew~J.}~\lnm{West}\orcid{0000-0002-0631-2393}}

\address[id=SwRI]{Southwest Research Institute, Boulder, Colorado, USA}
\address[id=ROB]{Solar-Terrestrial Centre of Excellence--SIDC, Royal Observatory of Belgium, Brussels}
\address[id=NU]{Department of Mathematics, Physics and Electrical Engineering, Northumbria University, Newcastle Upon Tyne, UK}
\address[id=ESAC]{European Space Agency, European Space Astronomy Centre, Madrid, Spain}
\address[id=NCEI]{National Centers for Environmental Information, National Oceanic and Atmospheric Administration, Boulder, Colorado, USA}

\runningauthor{D.B.~Seaton et al.}
\runningtitle{The SWAP Filter for EUV Solar Images}

\begin{abstract}
We present the SWAP Filter: an azimuthally varying, radial normalizing filter specifically developed for EUV images of the solar corona, named for the \textit{Sun Watcher with Active Pixels and Image Processing} (SWAP) instrument on the Project for On-Board Autonomy 2 spacecraft. We discuss the origins of our technique, its implementation and key user-configurable parameters, and highlight its effects on data via a series of examples. We discuss the filter's strengths in a data environment in which wide field-of-view observations that specifically target the low signal-to-noise middle corona are newly available and expected to grow in the coming years.
\end{abstract}
\end{opening}
%-------------------------------------------------

\section{Introduction}
     \label{sec:intro} 

A problem of understated difficulty and importance for observational coronal physics is simply finding ways to display and analyze observations so they are actually useful. The need for such techniques stems from a number of sources, but it generally boils down to a few key facts: the corona has a tremendous dynamic range in brightness, many interesting features and events are inherently faint and hard to detect, and various sources of noise can interfere with both imaging itself and can be amplified by image-processing techniques. Noise sources include instrumental effects and photon shot noise embedded in the signal itself. 

This problem is not unique to observations in the extreme ultraviolet (EUV), but it has emerged as an especially important one in this spectral band, particularly in light of the development of new EUV imagers -- and new ways of using these imagers -- that can observe the corona to heights as large as $\approx$6\,R$_{\odot}$. Such imagers include the \textit{Sun Watcher with Active Pixels and Image Processing} (SWAP) on the Project for On-Board Autonomy 2 (PROBA2) spacecraft \citep{2013SoPh..286...43S, 2013SoPh..286...67H}, the \textit{Extreme-Ultraviolet Imager} on Solar Orbiter (\citealt{Rochus2020, auchere2023_EUI}; see also the observations of a prominence eruption to $>6\,\mathrm{R}_{\odot}$ reported by \citealt{Mierla2022}), and the \textit{Solar Ultraviolet Imager} (SUVI) on NOAA's GOES-R line of spacecraft \citep{SEATON2020219, Darnel2022}. Both SWAP and SUVI have fields of view (FOV) that can reveal features in the EUV corona to heights greater than 2\,R$_{\odot}$, and both have been used in off-pointed campaigns that revealed EUV corona structures to substantially larger heights, as great as $\approx5\,\mathrm{R}_{\odot}$ \citep[e.g.][]{2014ApJ...781..100G, Ohara2019, Tadikonda2019, Seaton2021}. Observing the region between 1.5\,\--\,6\,R$_{\odot}$ -- the \textit{middle corona} \citep{West2022} -- is important for a wide variety of scientific questions including coronal mass ejection (CME) initiation, solar wind acceleration, and characterizing the processes that give the corona its large-scale structure.

Upcoming and proposed instruments, including the \textit{EUV CME and Coronal Connectivity Observatory} \citep[ECCCO:][]{Golub2020} and the \textit{Sun's Coronal Eruption Tracker} CubeSat \citep[SunCET;][]{Mason2021, Mason2022}, should reveal the EUV middle corona in dramatically more detail than has been possible, even with SWAP and SUVI. All the new observations that these instruments yield point to the great need for image-processing techniques that can help improve the accessibility of interesting features and events buried in these datasets.

In fact, this need is also shared by other data sets, including both those from the \textit{Association of Spacecraft for Polarimetric and Imaging Investigation of the Corona} of the Sun (ASPIICS) coronagraph \citep{2018A&A...612A..82S, 2019A&A...622A.101S} on the formation-flying PROBA3 mission and ground-based eclipse observations \citep{Habbal2007, Habbal2011} This and other new instruments will also observe the corona in largely the same region, and will likely encounter many of the same challenges, albeit in other wavelength regimes.

In this article we briefly describe our approach for processing these images to manage dynamic range and reveal features that would otherwise be invisible, which we refer to as the \textit{SWAP Filter}, thanks to its origins. These techniques have been demonstrated extensively on EUV coronal observations, including observations of large-scale EUV structure and evolution \citep{Seaton2013}, eruptive flares \citep{Seaton2018}, and off-pointed observations \citep{Seaton2021}, but they have not yet been described in detail in the literature.

In addition to the technique described here, there are many other resources for similar or related image processing available, most of which are included in common software packages such as \textsf{SolarSoft IDL}, the \textsf{SunPy} package \textsf{Sunkit-Image}, \textsf{jHelioviewer}, or other open-source repositories. These include Normalizing-Radial-Graded Filtering \citep[NRGF:][]{Morgan2006}, Fourier-normalizing Radial Gradient Filtering \citep[FNRGF;][]{Druckmullerova2011}, Noise Adaptive Fuzzy Equalization \citep[NAFE:][]{Druckmuller2013}, Multiscale Gaussian Normalizaton \citep[MGN:][]{Morgan2014}, Wavelet-Optimized Whitening \citep[WOW:][]{auchere2023_WOW}, \textsf{jHelioviewer's} $r^3$ radial brightness scaling \citep{Mueller2017}, and numerous other simpler techniques \citep[e.g.][]{Patel2022}. 

Although today this problem is typically addressed using software, techniques to manage dynamic range in coronal imaging have their origins many decades ago, during the era of photographic observations. One of the earliest examples was developed by Gordon Newkirk, who found that photographic images lacked sufficient dynamic range to capture the detail in the corona visible by eye during a total solar eclipse. In response, Newkirk devised an optical radially graded neutral-density filter that successfully compensated for the steep brightness gradient of the corona \citep{Eddy1989}. Newkirk first deployed his filter at an eclipse in 1966 in Bolivia, generating a stunning image of the corona that arguably revealed more detail than was visible to eye \citep{1967ARA&A...5..213N}. Subsequently, Newkirk-style filters were widely used for eclipse photography before digital imaging made it possible to assemble high-dynamic range (HDR) composites using multiple exposure times instead of radial-normalizing filters.

Although the origins of our technique are firmly rooted in the era of space-based electronic imaging, the fundamental motivation is the same as Newkirk's: during our work with SWAP images, we realized that there was more to see in our images than we could display using simple techniques. In this article, we describe the problem and the simple observation that led to the development of our approach (Section~\ref{sec:origins}), describe the approach itself and some of its configurable parameters (Section~\ref{sec:implementation}), and finally highlight a few examples of its applications (Section~\ref{sec:applications}) before making a few concluding remarks (Section~\ref{sec:conclusions}).

\section{The Origins of Our Filter Technique}
\label{sec:origins}
Early in the SWAP mission, we realized that SWAP's large FOV might reveal features and evolution in the solar corona that had not been well observed previously in these wavelengths. We began to produce movies composed of deep-exposure images spanning full Carrington rotations, using median stacking -- that is, computing the median of each image pixel in the temporal direction -- of many individual SWAP observations to suppress noise and enhance the visibility of faint features in the middle corona \citep[see Section~2.2 in][]{WestSwap2022}. We quickly realized, however, that without a technique to equalize the dynamic range in the corona, there was no path to take full advantage of these new observations, since we could not clearly visualize the large-scale features in the data.

\begin{figure}
\centering
\includegraphics[width=0.32\textwidth]{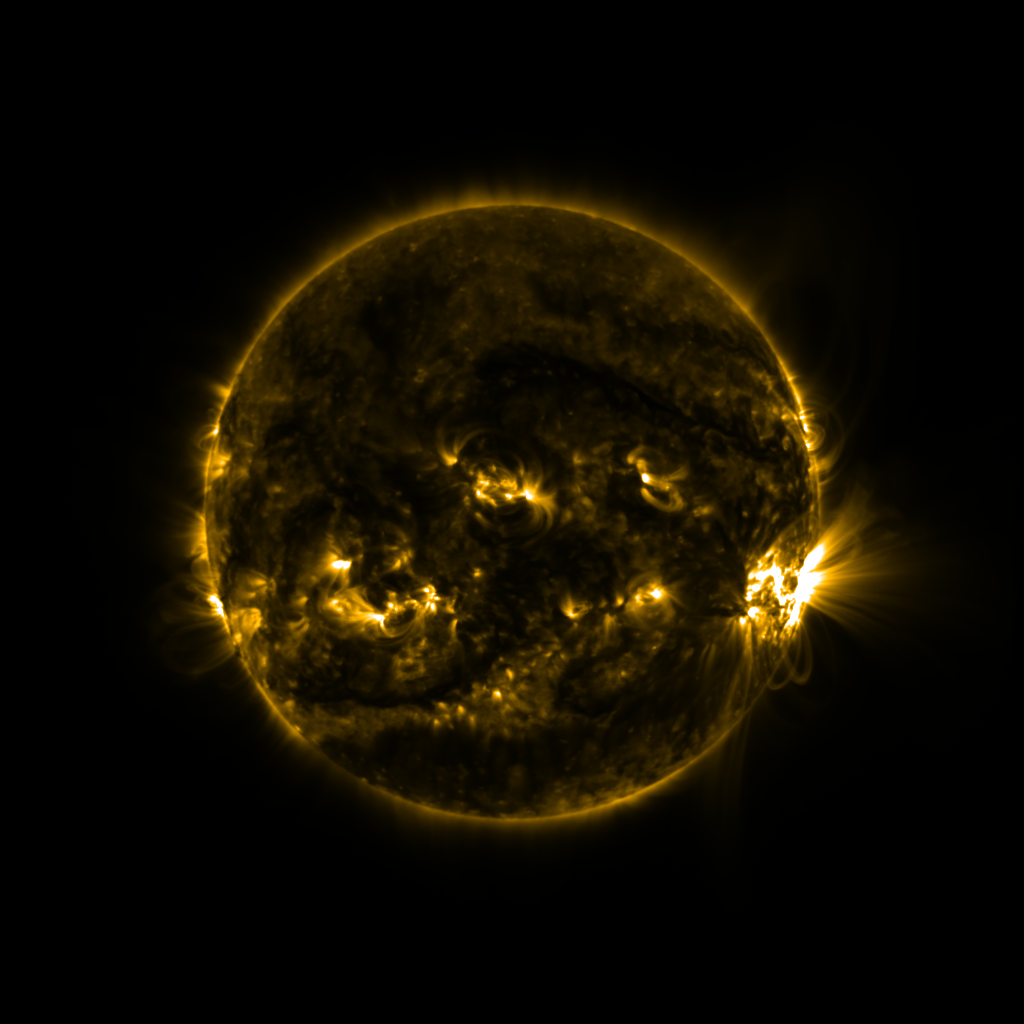}
\includegraphics[width=0.32\textwidth]{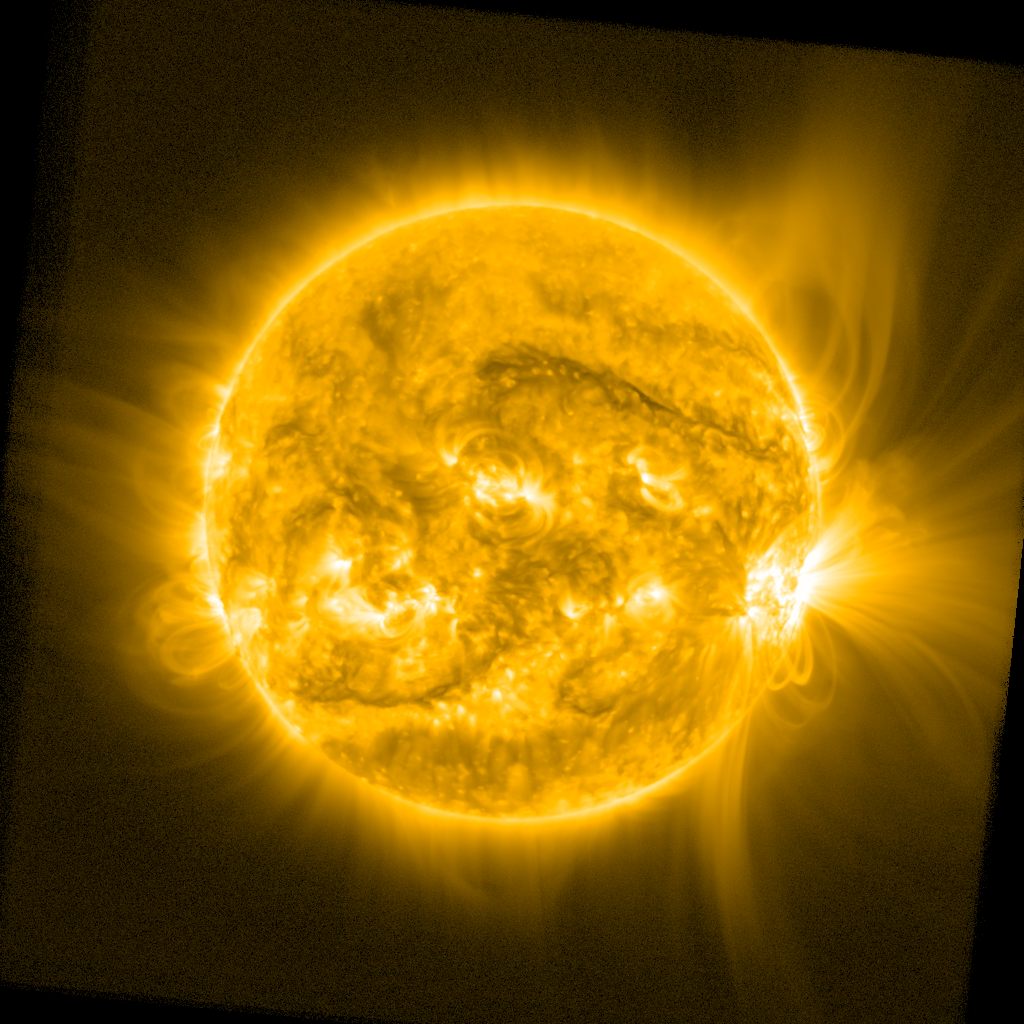}
\includegraphics[width=0.32\textwidth]{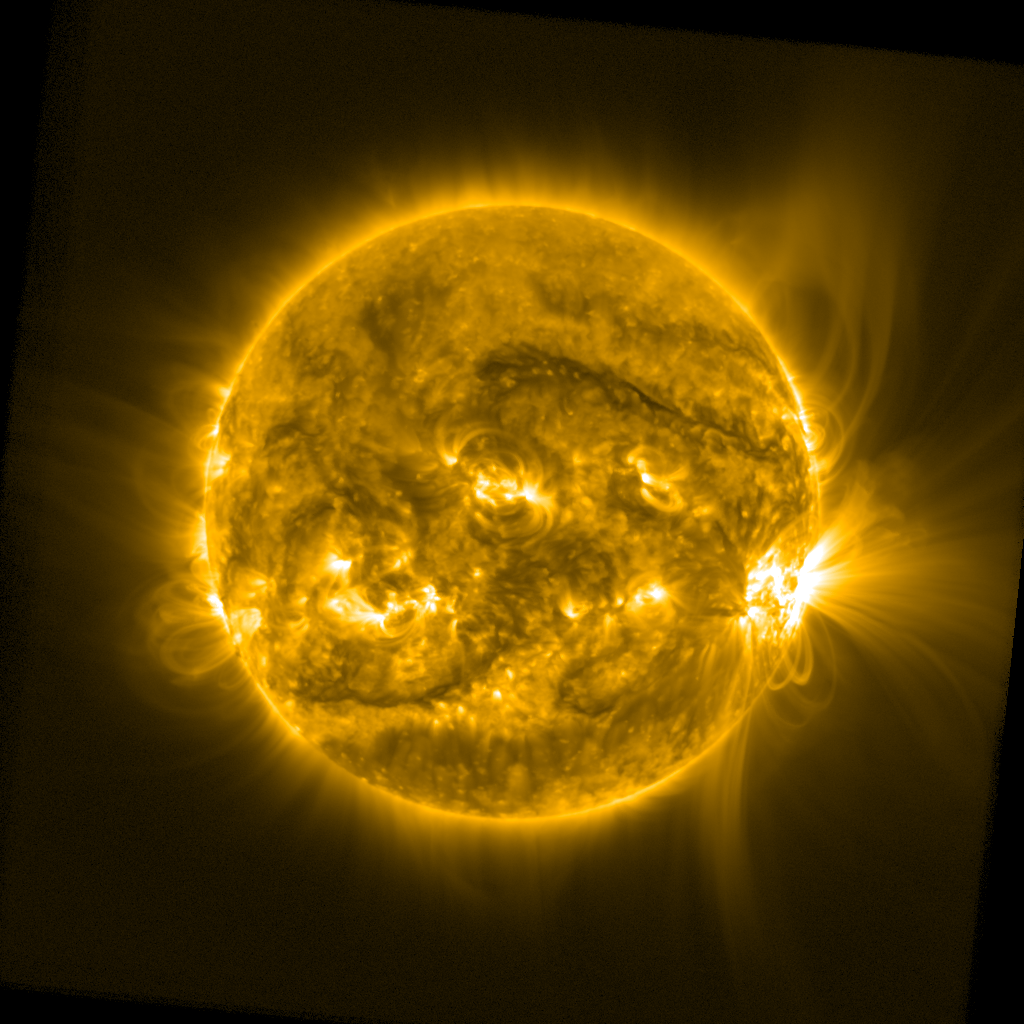}
\caption{The same SWAP image, displayed on a linear scale (\textit{left}), log scale \textit({center}), and \sfrac{1}{4} power \textit({right}). }\label{fig:scaling}
\end{figure}

Figure~\ref{fig:scaling} highlights this dilemma, showing an example of a deep-stacked SWAP image from 28 Oct 2014 at 13:21:13~UT, displayed using three different data-scaling approaches. The left panel uses a linear scale, which results in an image with too much contrast to be shown on typical displays. A common approach to address this problem is to use a logarithmic scaling (center), but this tends to suppress fine variations too much to visualize anything other than the very large-scale features. A third approach is to raise the image values to a power or other nonlinear function, \citep[e.g.][]{Lupton2004}, which reduces dynamic range enough to reasonably display the image without washing out fine features. We found a good value for SWAP observations was the \sfrac{1}{4} power, but found it was still not adequate to provide a detailed view of the entire corona within the FOV. Thus, while these various data scaling approaches are used widely for narrow-field EUV image data, we found that none were adequate to capture both large-scale structure, which can vary in brightness by up to five orders of magnitude over the FOV, and fine features, which often represent variations of $<$1\%.

A straightforward approach to overcoming this challenge is to rescale the data using a symmetrical radial gradient normalizing filter. Much like Newkirk's analog implementation, these filters are used to neutralize the steep radial falloff of coronal brightness as a function of height above the disk. Typically one generates the filter by sampling the coronal brightness in successive rings at greater heights, and then dividing the coronal image by the resulting array to eliminate the gradient with height, leaving only local variations to be displayed\footnote{An implementation of this type of filter is available in the \textsf{SunPy} Examples Gallery: \url{https://docs.sunpy.org/en/stable/generated/gallery/computer\_vision\_techniques/off\_limb_enhance.html\#sphx-glr-generated-gallery-computer-vision-techniques-off-limb-enhance-py}.}. 

Although filters like this have proven highly successful for visible-light coronal imaging, we found that this simple approach was not satisfactory for large-scale EUV images due to the large variation in brightness between polar coronal holes and equatorial streamers and other bright features.

We sought another approach to ``normalize'' -- that is, divide -- the data that could be adapted to adjust the degree of data scaling as a function of azimuth around the Sun. (Note that throughout this article we adopt the term normalize to refer to a process in which data are rescaled by division by a function or array.) Because at the time we were working with observations that spanned complete Carrington rotations, we had a large amount of data that revealed the evolution of the Sun and corona over a full solar rotation -- or even longer -- readily available to work with.

As an initial experiment, we adopted a concept similar to the monthly minimum images \citep[e.g. Section 10 in][]{Morrill2006}, which are used to subtract background (stray light and F-corona) from coronagraph observations, to see whether a similar type of stacking might yield an image that we could use to normalize our data. This approach turned out to be very effective, because long-term averages smooth out most (but not all) local variations and leave a largely smooth -- but azimuthally variable -- fall-off in intensity that can serve as the basis to normalize the height gradient of an image. 

Because the large-scale features in the off-disk corona often persist for many months \citep{Seaton2013}, long-term image stacks preserve some of the variation of coronal brightness as a function of latitude, which leaves a normalizing filter that acts differently at equatorial latitudes, where bright active regions are often present; at mid-latitudes, where extended coronal fans are often observed; and at the poles, where coronal holes are most common. In our case, we found median-stacking the data to be most effective approach to generating these filters, since the median suppresses both temporal noise and transient events that might otherwise skew the behavior of the filter.

\begin{figure}
\centering
\includegraphics[width=0.49\textwidth]{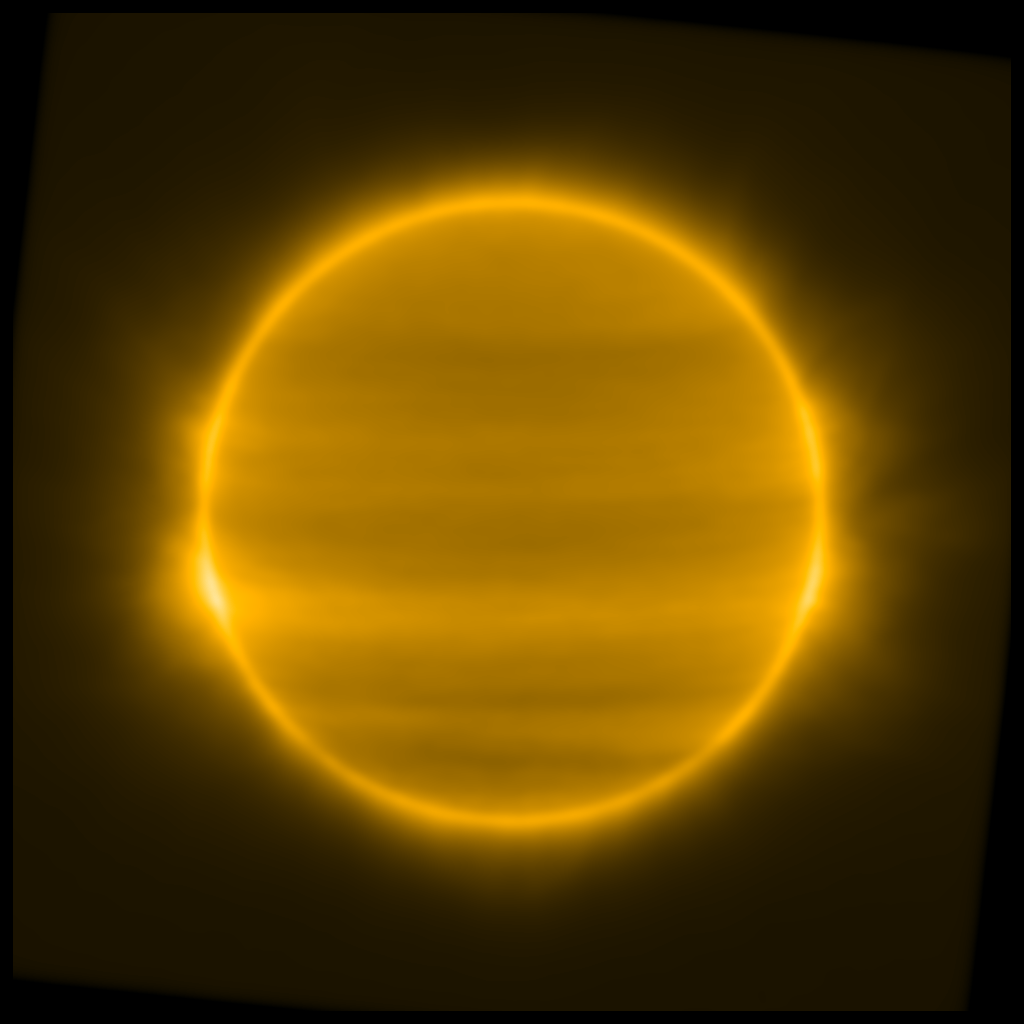}
\includegraphics[width=0.49\textwidth]{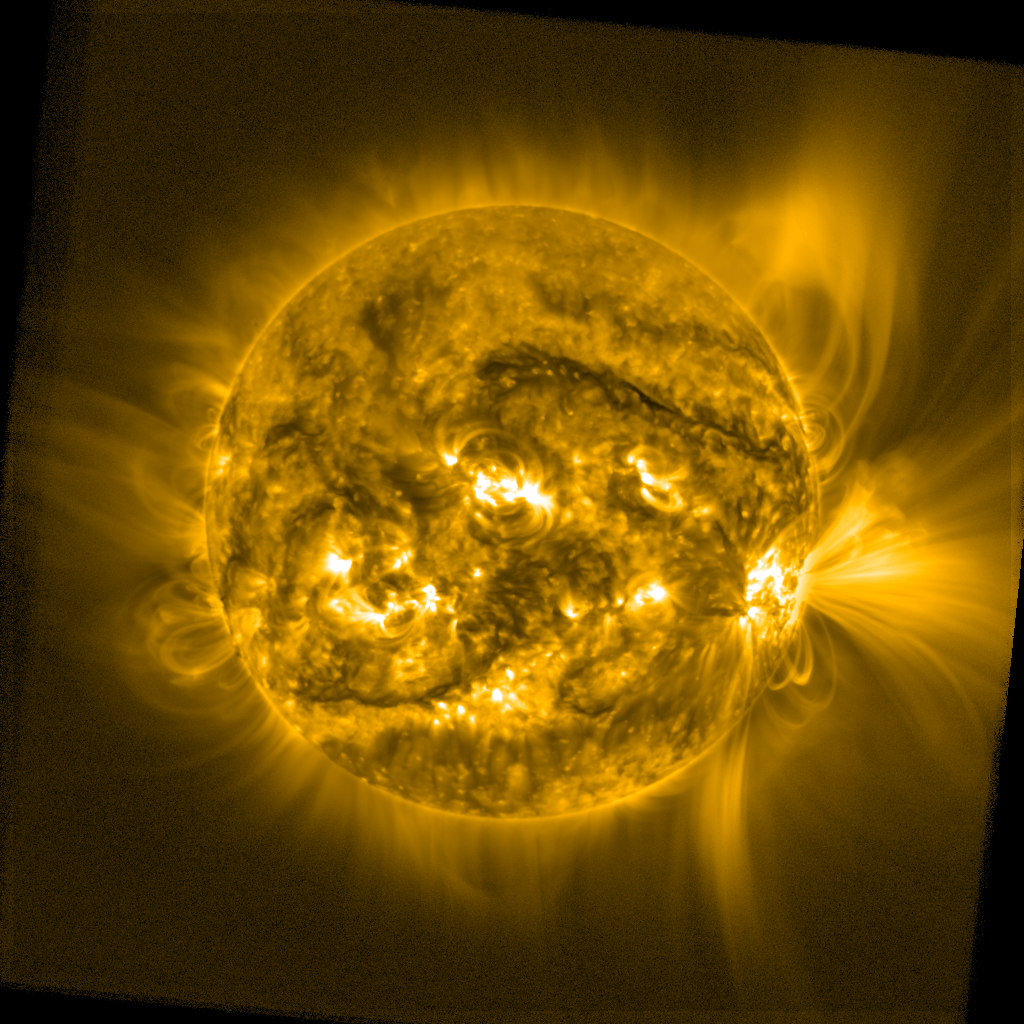}
\caption{ (\textit{left}) Running median of all SWAP images during Carrington Rotation 2156 displayed at \sfrac{1}{4} power and (\textit{right}) the result of using this running median as an image filter to normalize the same test image as in Figure 1 (28 Oct 2014 13:21:13~UT). An accompanying animation in the Electronic Supplementary Material shows the results of applying this type of filter to a full solar rotation's worth of observations.}\label{fig:median}
\end{figure}

Figure~\ref{fig:median} shows an example of a long-term median-stack image generated using all SWAP observations during Carrington Rotation 2156, and a sample of an image (the same image shown in Figure~\ref{fig:scaling}) processed using this image as the basis for a normalizing filter. In addition to their value in equalizing the brightness in the off-disk region, an important feature of these filters is the bright bands left by active regions as they transit the solar disk. When used to normalize a full-Sun image, these bands suppress a bit of brightness at latitudes where the most prominent active regions appear, and comparatively amplify quiet-Sun features. The net result is a disk image that is normalized so both bright active regions and dark coronal holes and filaments are well scaled and overall contrast is better than what can be achieved with simpler image normalization techniques. It is important to note that the filtered images should not be used to determine absolute brightness or density on their own, but they are excellent for determining where signal is coming from, and characterizing coronal morphology. 

Because our experimental strategy leveraged all of the data for a full solar rotation, it turned out that these filters were extremely useful in processing movies of the evolution of the extended EUV corona over one or more solar rotations. (See the animation accompanying Figure~\ref{fig:median} for an example.) However, although this experimental approach yielded dramatic results, it suffered from an important drawback: because the method requires a full Carrington rotation of SWAP observations to generate the filter, it was only practical for processing SWAP's long-duration movies, when the data were already available. Thus we sought an alternative approach that could be more easily deployed for short-duration events such as CMEs and solar flares, but would share most of the benefits that we realized from this technique.

The Carrington-rotation-filter method yielded filters with several specific, helpful features: First, they capture the bright limb and the gradual falloff in brightness with increasing height, which largely determines the overall dynamic range of the image. Second, they capture gradual changes as a function of azimuthal angle, allowing the filter to adapt to appropriately normalize both bright extended features at lower latitudes and fainter features in the coronal-hole regions near the poles appropriately. Finally, they achieve good normalization across the disk and off-limb regions, specifically improving contrast on the disk because they include band-like features that result from the passage of either active regions or darker features, such as coronal holes and filaments, at different latitudes. In the Section~\ref{sec:implementation} we describe how we translated these naturally derived features into a filtering approach that requires only a single input image and subsequently applied this filter technique to a variety of observations from SWAP and other instruments.

\section{Implementation and Use}
\label{sec:implementation}
Our strategy uses the three key features that we identified in filters based on Carrington-rotation image stacks discussed above: it captures the fall-off in brightness above solar limb, it captures the slow variation of this gradient as a function of azimuth, and captures the latitudinal distribution of brightness on the solar disk. The method to generate filter arrays is implemented in a \textsf{SolarSoft IDL} function, called \textsf{p2sw\_image\_filter.pro} in the PROBA2/SWAP software package.

The generation of the filter is mostly implemented in two straightforward steps, both of which leverage transformations of coordinates that facilitate data smoothing and averaging in directions that are not practical in image coordinates. 

We generate the \textit{off-disk} part of the filter by reprojecting the part of the image above the limb into polar coordinates, with azimuthal angle along the horizontal axis and radial height on the vertical. By smoothing the image in the azimuthal direction with a spatial median filter, which does not affect the radial direction, we generate an array that varies slowly in azimuth, but captures the remaining local gradient with good fidelity. We pad the edges of the transformed image with copies of itself to ensure that the processing wraps smoothly around the Sun at the discontinuity in the azimuthal angle (i.e. $0^{\circ}/360^{\circ}$). After the smoothing process, we transform the processed array back from radial coordinates to Cartesian image coordinates. 

We generate the \textit{on-disk} part of the filter by transforming the disk image to a simple cylindrical map projection, which accounts for the observatory's viewing angle in heliographic latitude (that is, the \textit{B$_0$}-angle). We then compute the median value for each latitude (that is, across the longitudinal direction) and replace all the pixels in the map with their appropriate median. Again, we transform this back to image coordinates.

These two processed parts of the filter are merged into a single image and modulated using a set of user-configurable parameters (described below), yielding a basis image to perform radial filtering that shares its broad characteristics with the Carrington-rotation-stack images that gave us the idea for this approach.

\begin{figure}
\centering
\includegraphics[width=0.49\textwidth]{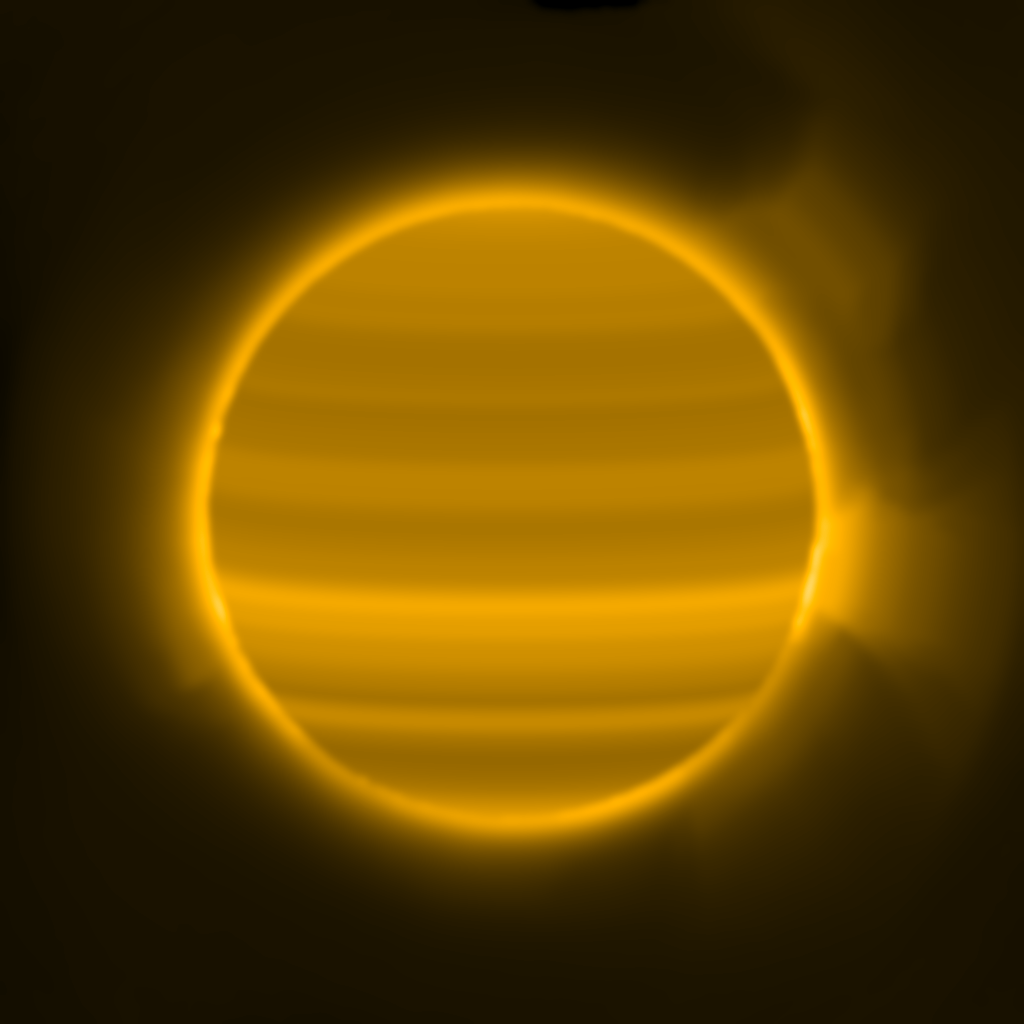}
\includegraphics[width=0.49\textwidth]{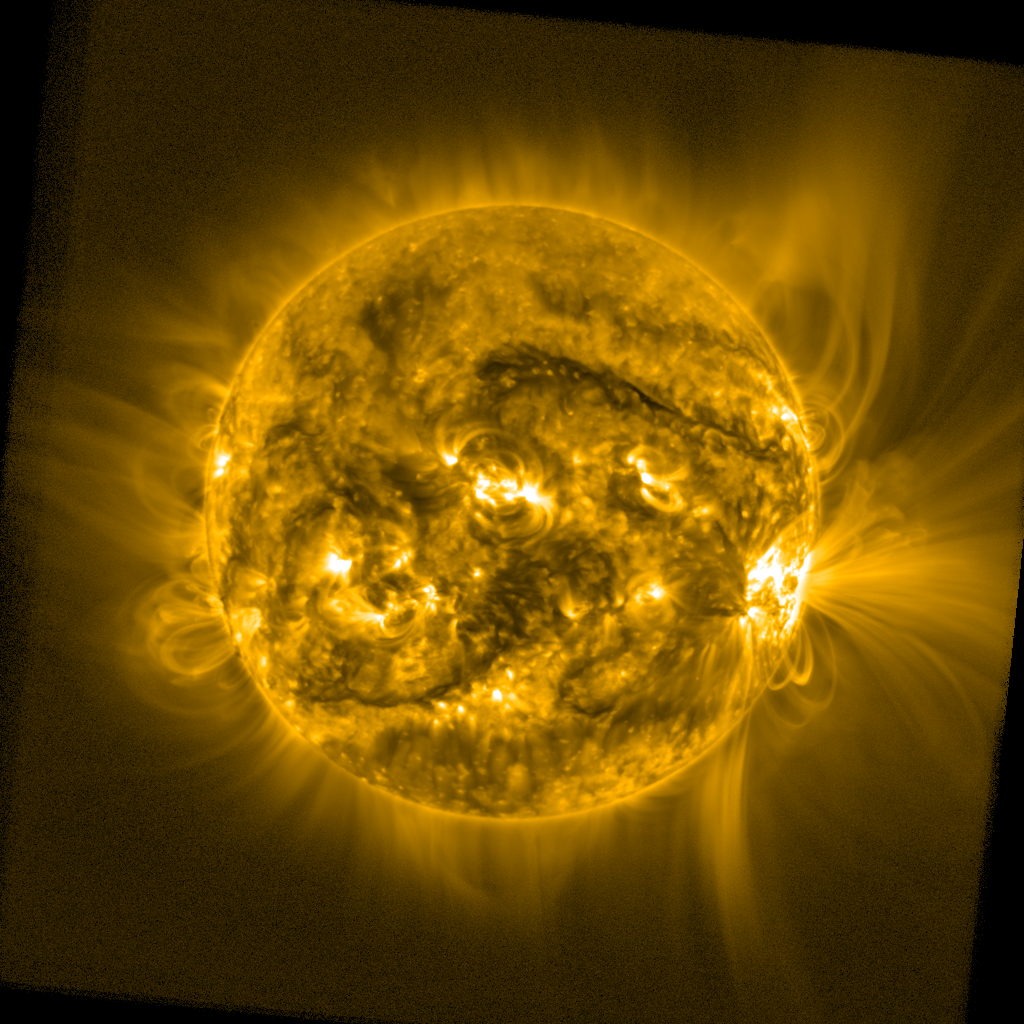}
\caption{Filter image generated by our method displayed at \sfrac{1}{4} power (\textit{left}) and the result of using this filter on our test image from 28 Oct 2014 13:21:13~UT (\textit{right}). Though specific features differ somewhat between this and the processed image in Figure~\ref{fig:median}, the overall results are comparable without the requirement of large quantities of data.}\label{fig:filter_example}
\end{figure}

Figure~\ref{fig:filter_example} shows the resulting filter and processed image, displayed using the same parameters as Figure~\ref{fig:median}. The method that we describe here is more sensitive to the specific features visible in the image used to manufacture the filter than the long-term stacking approach is, and thus more strongly modulates the characteristics of these specific features. The resulting image has overall less contrast than the processed image in Figure~\ref{fig:median}, but in each case both the overall global structure of the corona and its local-scale variations, both on- and off-disk, are far more visible than in any of the images in Figure~\ref{fig:scaling}.

The filter's overall sensitivity to local features can be adjusted somewhat by altering the width of the median filter used off-disk, which is one of several user-configurable parameters in the code. Here we used a value of $\pm15^{\circ}$, which we found matches the scale of many typical off-disk coronal features, but which can be adjusted to emphasize different scales. Very large values diminish the filter's ability to adapt azimuthally to changing coronal brightness, while very small values make the filter too sensitive to local variations and suppress all but the most fine-scale features. The filter can also be tuned by an overall smoothing parameter, which is useful both for setting the scale of features to emphasize and suppressing small artifacts that sometimes develop during the filter generation. Here we used a 2D Gaussian blur with a $\sigma$ value of four pixels on the final filter in Cartesian space, but both the blurring method and width are adjustable depending on the user's needs.

There are two other very important parameters that must be set to obtain a high-quality, low-noise, processed image that retains the overall natural appearance of typical solar EUV images: a bright limb, bright active regions, and a gradual decrease in brightness with height. (Such aesthetic choices are not necessarily requirements of a generic image filter; they help meet one of the objectives of our specific filter: namely to generate images and movies for public-outreach materials.) The filter, as derived, acts uniformly across the entire image. If applied with no adjustments, the brightest features will be re-normalized to have the same brightness as the faintest features, while the faintest features, which are also the noisiest in most images, will be enhanced to match the brightness of all other features. The result is an image that both lacks contrast -- or is excessively ``flat'' in appearance -- and excessively noisy in its faintest regions. We therefore must tune the filter slightly to avoid both of these outcomes.

This is straightforwardly achieved by modulating the overall filter array $\left[\mathbf{F}\right]$ to yield a final filter $\left[\mathbf{F'}\right]$, in the following way,
\begin{equation}
\mathbf{F'} = \left( \mathbf{F} + t_0 \right)^{c_0}.
\label{eq:filt_mod}
\end{equation}

Here, $t_0$ is a constant offset, which we refer to as the \textit{tapering parameter}, that depends on the image background brightness (or can be adjusted by the user) and reduces the amount by which the filter enhances the faintest pixels -- and specifically the noise in these regions. $c_0$ is a factor that we refer to as the \textit{crush factor} and is defined such that $0 < c_0 < 1$. This factor reduces the overall normalization effect of the filter, particularly on bright values. Careful tuning of these parameters can significantly improve the overall image result, but any filter that amplifies the faintest part of the image without also applying noise reduction is likely to also enhance noise in faint regions as well, thus there is always a trade-off to make between how aggressively to filter regions where the signal-to-noise ratio is low.

For the SWAP images in both Figures~\ref{fig:median} and \ref{fig:filter_example}, we set the constants in Equation~\ref{eq:filt_mod} to be $t_0 = 1.5$ and $c_0 = 0.75$. In our experience, the distribution of brightness in typical wide-FOV EUV images is such that the value of $t_0$ should be roughly the median of all pixels in the filter image, but manual fine-tuning is often required to obtain the user's desired appearance for the image. Note that the value of $t_0$ differs between the two techniques that we illustrate in this article, so we select $t_0 = 1.5$, which is larger than the median of the Carrington-rotation-stack and smaller than the median of our method's filter, for both cases to ensure the overall image is treated roughly equally by both methods.

\begin{figure}
\centering
\includegraphics[width=1.0\textwidth]{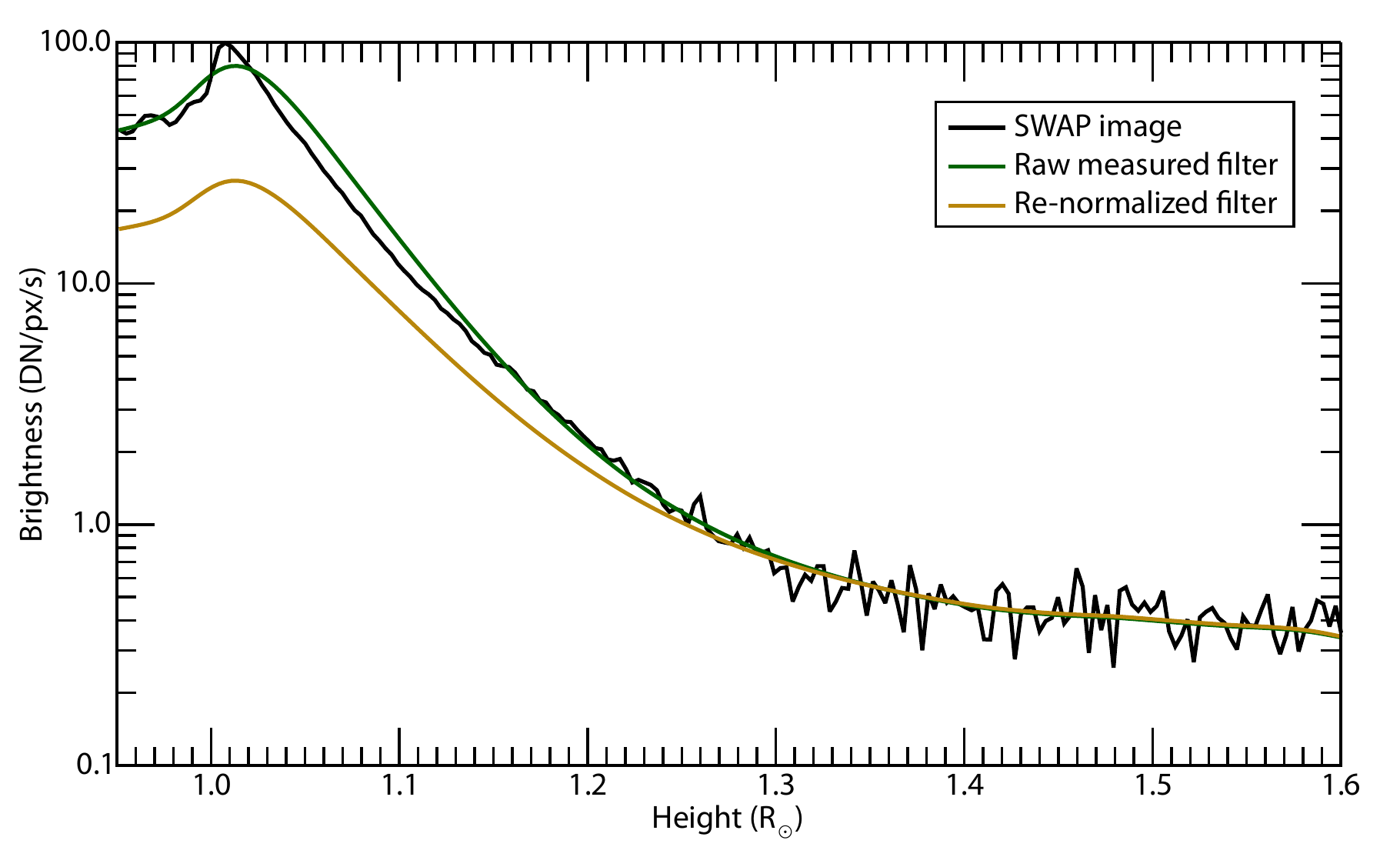}
\caption{Brightness of a cross-section of our test image from South Pole to the edge of the FOV compared to the unprocessed filter in Figure~\ref{fig:filter_example} (\textit{green}) and the same rescaled using $c_0 = 0.75$ (\textit{gold}).}\label{fig:cross_sec}
\end{figure}

Figure~\ref{fig:cross_sec} illustrates why it is necessary to rescale the filter using $c_0$ before processing an image using some cross-sectional cuts through the data in our sample image. If we do not compress (or ``crush'') the filter dynamic range, dividing by the filter will completely flatten all variation in brightness with height above limb, which results in an image that looks significantly modified from the original observations to many viewers. Crushing this function using $c_0=0.75$ has little effect on how the filter treats faint features, but reduces somewhat how much it flattens bright features, resulting in an image that retains some -- but much less -- of the gradient from limb to the large heights and is both appropriate for display and more natural in appearance.

\begin{figure}
\centering
\includegraphics[width=0.49\textwidth]{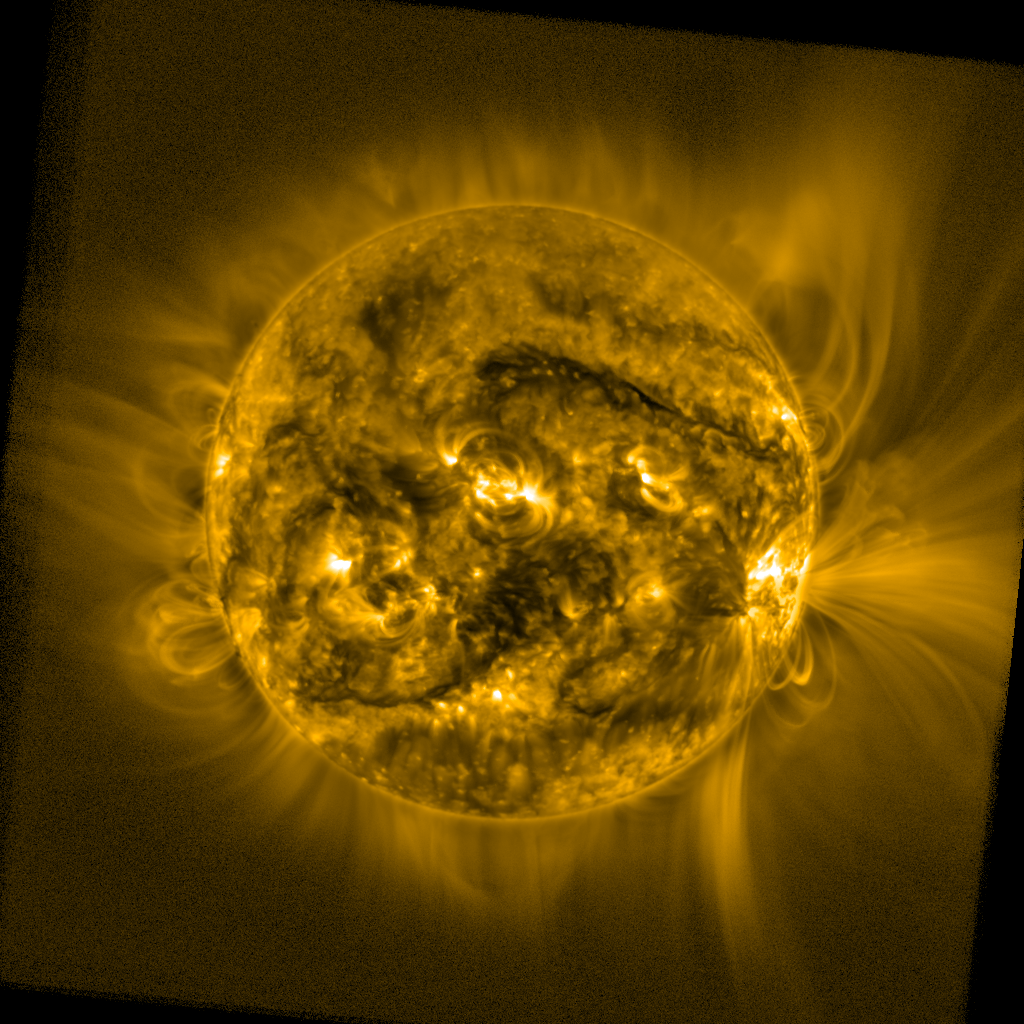}
\includegraphics[width=0.49\textwidth]{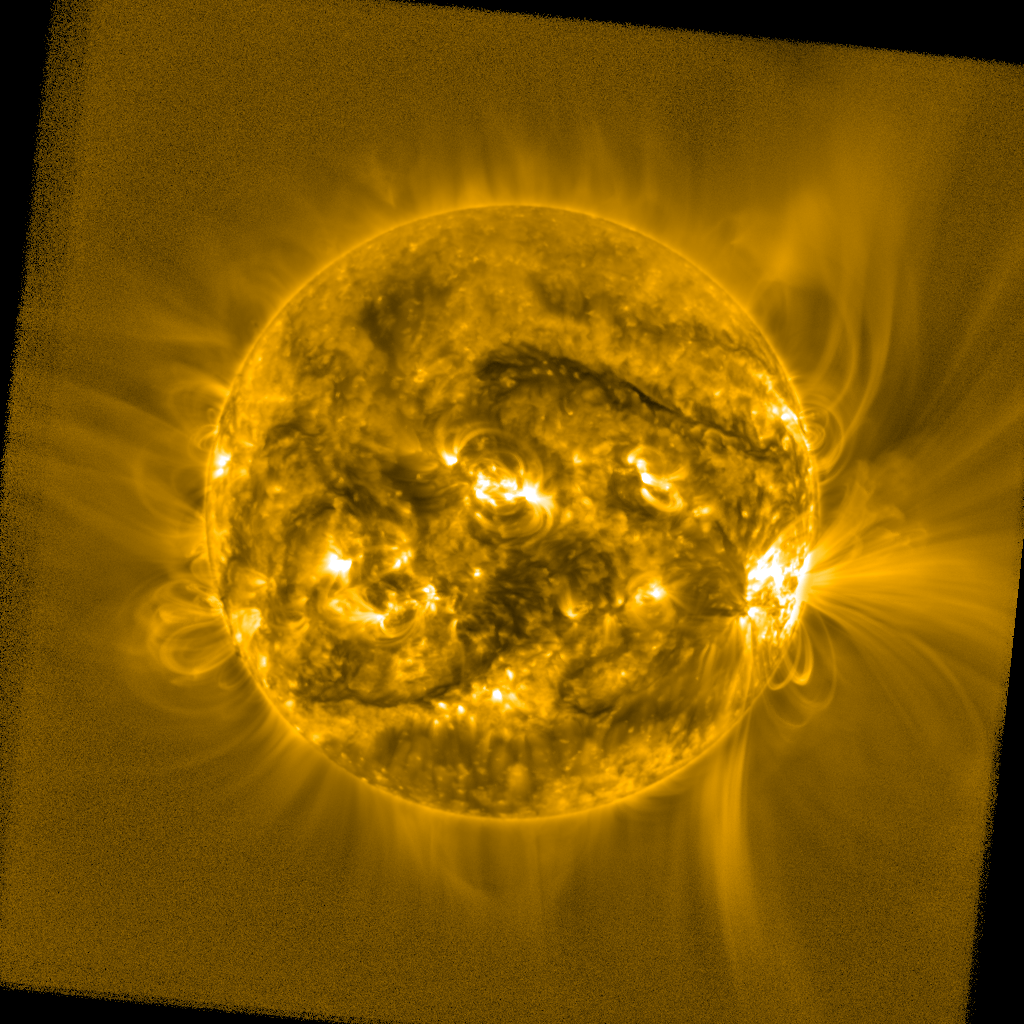} \\
\includegraphics[width=1.0\textwidth]{Figures/filtered_filter.png}
\caption{Comparison of a filtered image with $c_0=1.0$ and $t_0=1.0$ (\textit{top left}; i.e. no rescaling, but including low-end tapering) resulting in a ``flat'' looking result and with $c_0= 0.75$, but $t_0 = 0.0$ (\textit{top right}; rescaling, but no low-end tapering), which leads to an over-enhancement of background brightness and low-level temporal noise, and (\textit{bottom}) the optimally processed image from Figure~\ref{fig:filter_example} (28 Oct 2014 13:21:13~UT).} \label{fig:image_factors}
\end{figure}

Figure~\ref{fig:image_factors} illustrates the results of neglecting these two parameters with two processed images. This yields images that either lack contrast (the $c_0 = 1.0$ case; left) or have excessive enhancement of the background (the $c_0 = 0.75$, $t_0 = 0.$ case; right). The overall result of appropriate selection of these parameters, as shown in Figure~\ref{fig:filter_example}, yields a final product that achieves balance between contrast, dynamic range, and resemblance to the original observation without excessively amplifying noise in dark regions.

Two final choices remain before the processed image can be displayed: how to scale the final filtered image for display, if at all, and how to set the appropriate limits to clip the displayed image's dynamic range. We found that our filtered images still have enough variation with height that it is useful to adjust the image scaling, and further, that scaling the image to the \sfrac{1}{4} power yields a subjectively aesthetically appealing result -- though, as we discuss in Section~\ref{sec:origins}, a variety of nonlinear functions can work depending on the specific need. For many of the processed images we have generated, we find that a good display range for the filtered and scaled image is between 0.5\,\--\,2.0 renormalized counts. We used this combination for the images in this article, but users are encouraged to experiment and find display approaches that suit their particular needs and data. Some examples in the \textsf{p2sw\_image\_filter.pro} code itself illustrate how to implement this and display an image in IDL. 

In the example above, the fully renormalized image $\mathbf{\left[I'\right]}$ is then described by
\begin{equation}
\mathbf{I'} = \left[ \left( \mathbf{I_0}/\mathbf{F'} \right)^{1/4} \right]^{2.0}_{0.5},
\label{eq:image}
\end{equation}
where, $\mathbf{I_0}$ is the unprocessed, calibrated image and $\mathbf{F'}$ is the filter generated in Equation~\ref{eq:filt_mod}. $\mathbf{I'}$ is re-scaled as a byte-array using the values indicated.

\section{Applications}
\label{sec:applications}

Although we have refined these filtering techniques somewhat since first developing them, largely to make them more robust to anomalies in image data, they have been used effectively on a variety of EUV image data and applications, including tracking the long-term evolution of large-scale features in SWAP data \citep{Seaton2013}, observations of eruptive solar flares in AIA \citep{Seaton2017} and SUVI \citep{Seaton2018, Veronig2018}, as well as the EUV waves that often accompany these events \citep{deKoning2022}, and even on simulated observations of eruptions \citep{Mason2022} from planned instruments \citep{Mason2021}.

\begin{figure}
\centering
\includegraphics[width=1.0\textwidth]{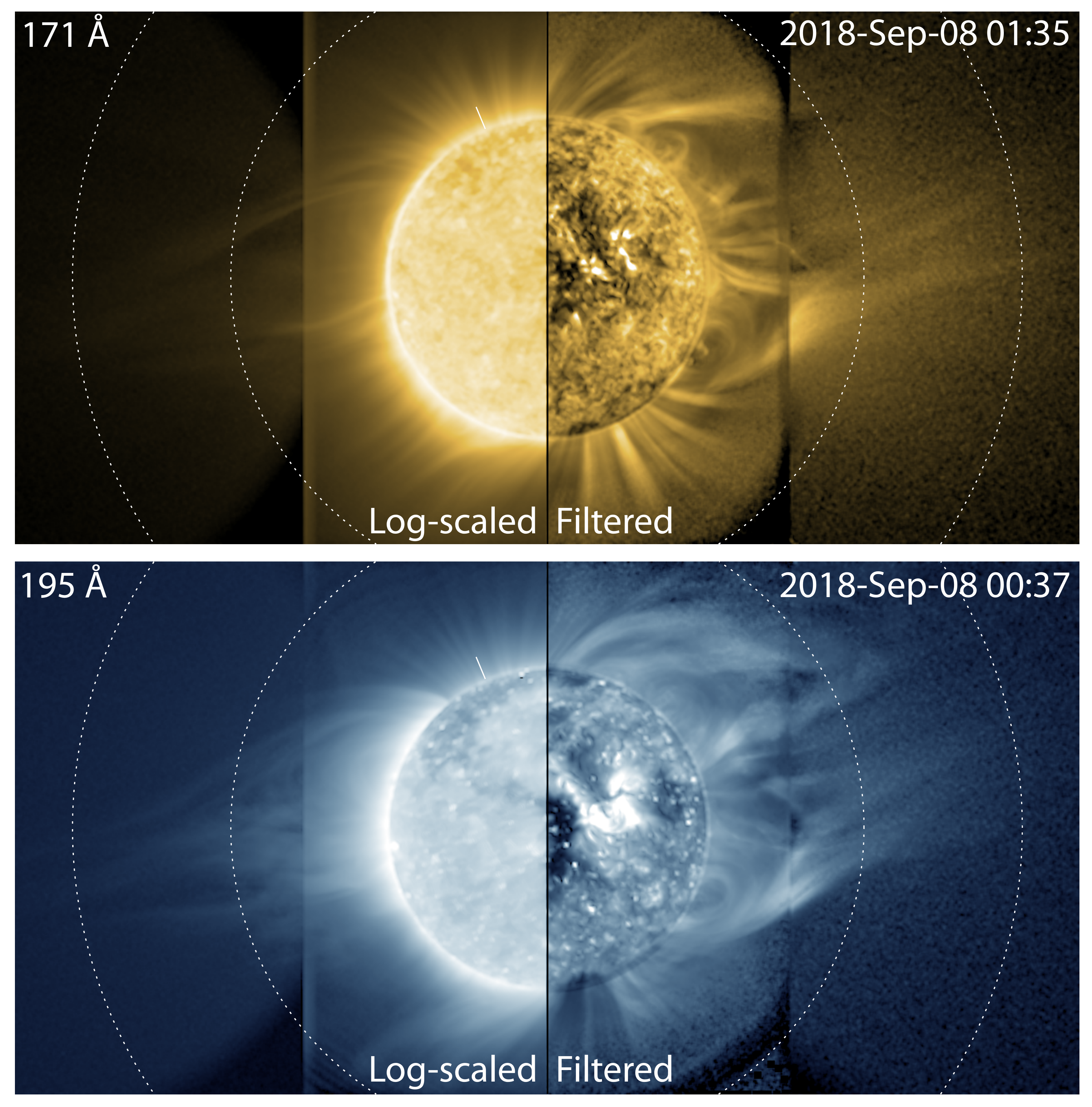}
\caption{Comparison between log-scaled and filtered data from a SUVI offpoint campaign in 2018. Concentric dotted circles indicate heights of 2 and 3~R$_\odot$. Solar North is indicated by the tick mark near the top of the solar disk. Adapted from \citet{Seaton2021}, used with permission.}\label{fig:suvi_comp}
\end{figure}

These techniques have proven especially valuable for ultra-wide-field observations of the EUV corona \citep{Tadikonda2019, Seaton2021}, where the dynamic range from limb to 3~R$_\odot$ can reach almost five orders of magnitude. Figure~\ref{fig:suvi_comp} contrasts the detailed structure visible in wide-field EUV images from SUVI when processed with our filter with the minimal structure visible when these images are displayed using log-scaling. Enhanced observations such as these make it possible to characterize the very faint dynamics associated with the origins of the solar wind's embedded structure \citep{Chitta2022}.

\begin{figure}
\centering
\includegraphics[width=1.0\textwidth]{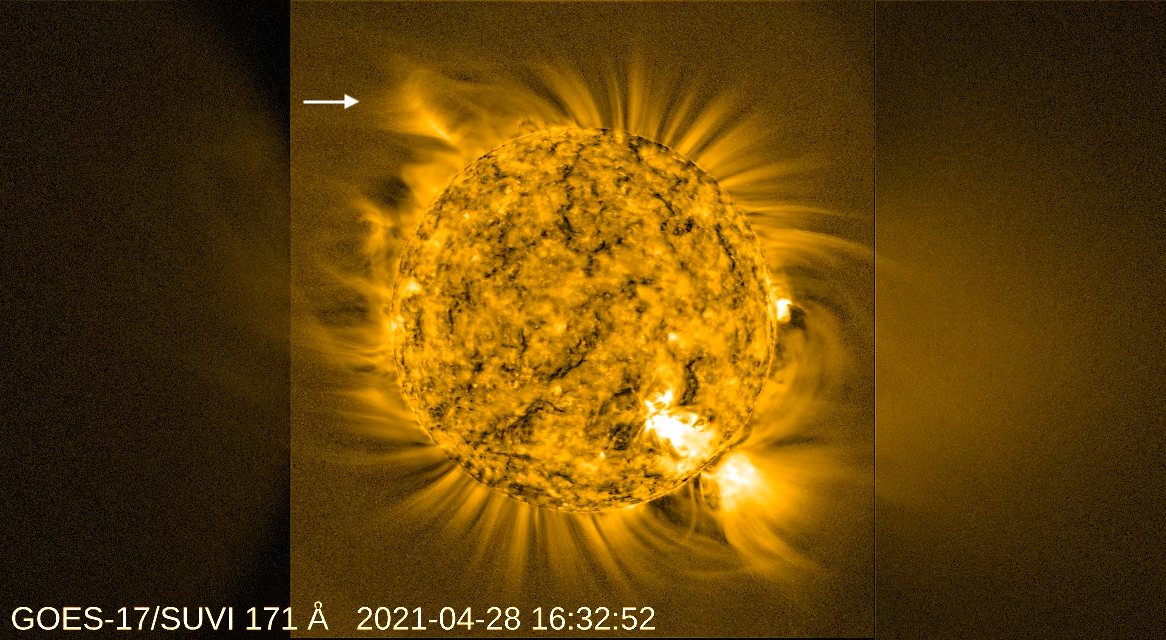}
\caption{SUVI image of extended EUV corona from a campaign during April 2021, showing an erupting prominence cavity, indicated by arrow near the northeast limb. See also the accompanying animation in the Electronic Supplementary Material.}\label{fig:suvi_eci}
\end{figure}

Figure~\ref{fig:suvi_eci} and the accompanying animation show how processing with our filter makes visible multi-scale dynamics in wide-field EUV observations. The movie makes visible both small, ubiquitous, jet-like features and several large eruptions near the east limb. These faint variations would not be detectable compared to the steep radial gradient in the brightness without this processing. \citep[See][for a complete discussion of the processing required for these SUVI mosaic images and the origins of artifacts in the resulting images and movies.]{Seaton2021}

\section{Comparison to Other Techniques }
\label{sec:comparison}

Since the development of the SWAP Filter, early in PROBA2's mission, a number of powerful, new, image-processing techniques have also been published \citep[e.g.][]{Druckmullerova2011, Druckmuller2013, Morgan2014, auchere2023_WOW}. The fundamental idea underpinning our approach was very simple: to adapt well-known radial-normalizing filter approaches -- dating all the way back to \citet{1967ARA&A...5..213N} -- to be more responsive to azimuthal variations, which are particularly pronounced in the EUV corona. The filter's primary purpose is thus to reduce dynamic range -- as a function of altitude within localized features off-disk, and as a function of latitude on-disk -- to improve display of data that span many order of magnitude in radiance.

Other methods filter specific spatial frequencies to emphasize and sharpen features with specific characteristics, using, for example, wavelets (e.g. \citealt{StenborgCobelli2003} and \citealt{auchere2023_WOW}) or multi-scale Gaussian normalization \citep{Morgan2014}. Although these techniques can be tuned to match features on many scales, they are particularly powerful for enhancing the visibility and clarity of low-contrast features on fine scales (see, for example, Figure~3 in \citet{auchere2023_WOW} or Figure~6 in \citet{Morgan2014}).

In contrast, the SWAP filter's strength lies in its ability to make visible large-scale coherent features without trading off localized contrast -- as with logarithmic or power scaling (compare Figure~\ref{fig:scaling} and Figure~\ref{fig:filter_example}). In this sense, the SWAP filter has much in common with other radial normalizing filters such as the NRGF and FNRGF, which are optimized specifically for off-disk image processing. The SWAP filter marries its off-disk processing technique to a related on-disk filtering technique, but is nonetheless fundamentally similar in approach to these methods.

\begin{figure}
\centering
\includegraphics[width=0.49\textwidth]{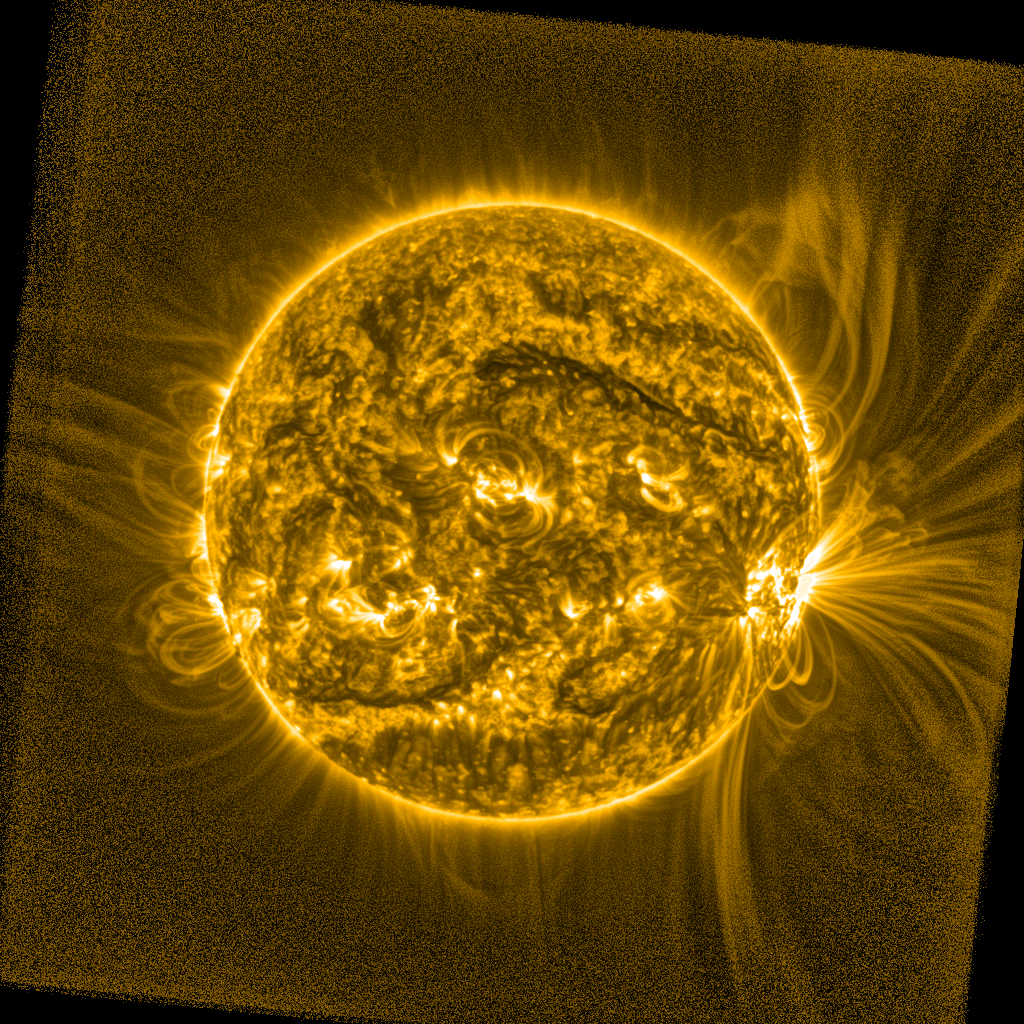}
\includegraphics[width=0.49\textwidth]{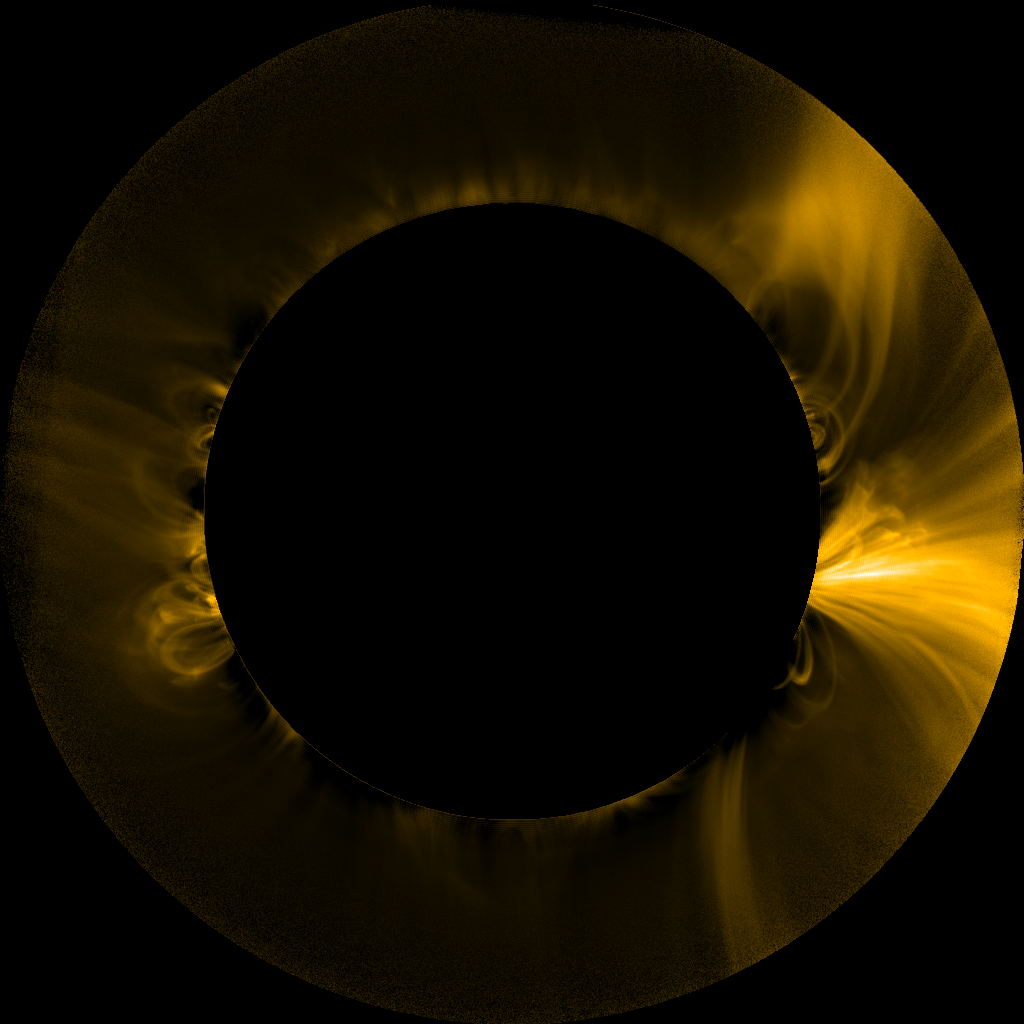}
\caption{(\textit{left}) MGN-filtered version and (\textit{right}) NRGF-filtered version of the image shown in Figure~\ref{fig:filter_example}, illustrating similarities and differences in the same image when processed with different techniques.}\label{fig:mgn_nrgf}
\end{figure}

Figure~\ref{fig:mgn_nrgf} presents examples of the same SWAP image shown elsewhere in this article processed with the MGN and NRGF filters, implemented in the \textsf{sunkit-image} Python package\footnote{See \url{https://docs.sunpy.org/projects/sunkit-image/en/latest/} for additional info and documentation.}. While the MGN-processed image clearly reveals coherent structures on multiple scales, the overall appearance of the image -- particularly off of the solar disk -- is somewhat different from the SWAP filter. The spatial filtering of the MGN suppresses the low-frequency background brightness in the image, strongly sharpening features on smaller scales and rendering them with largely uniform intensity across the full FOV. Like the SWAP filter, the processing of the disk image improves the visiblity of loops and large-scale features such as the filament in the northwest part of the Sun, but the MGN makes small-scale variations within these features much more prominent. Furthermore, the MGN can be flexibly tuned to both highlight features on specific scales, by adjusting the specific spatial scales isolated by the filter, and remix the relative importance of each frequency in the image, by adjusting weights for each individual spatial component in the final image. (Here we set the MGN to isolate a range of spatial scales ranging from just over a single pixel to 40 pixels, with the strongest weights for the large-scale features, but alternate scalings are easily achieved via configurable parameters.)

The NRGF image, on the other hand, more closely resembles the processed image in Figure~\ref{fig:filter_example}, because the approach used by the NRGF filter is similar to the off-disk approach used in the SWAP filter. A notable difference is that the NRGF processing -- which is described by \citet{Morgan2006} as being specifically developed to address the steep brightness gradient in coronagraph images -- is not optimized for use on disk. This region is often omitted from NRGF-processed images (or, in the case of coronagraphic images, is simply unavailable). Nonetheless, because both the SWAP filter and NRGF were both developed to address the radial gradient in coronal images, it is not surprising that they emphasize many of the same large-scale coronal features.

\begin{figure}
\centering
\includegraphics[width=0.99\textwidth]{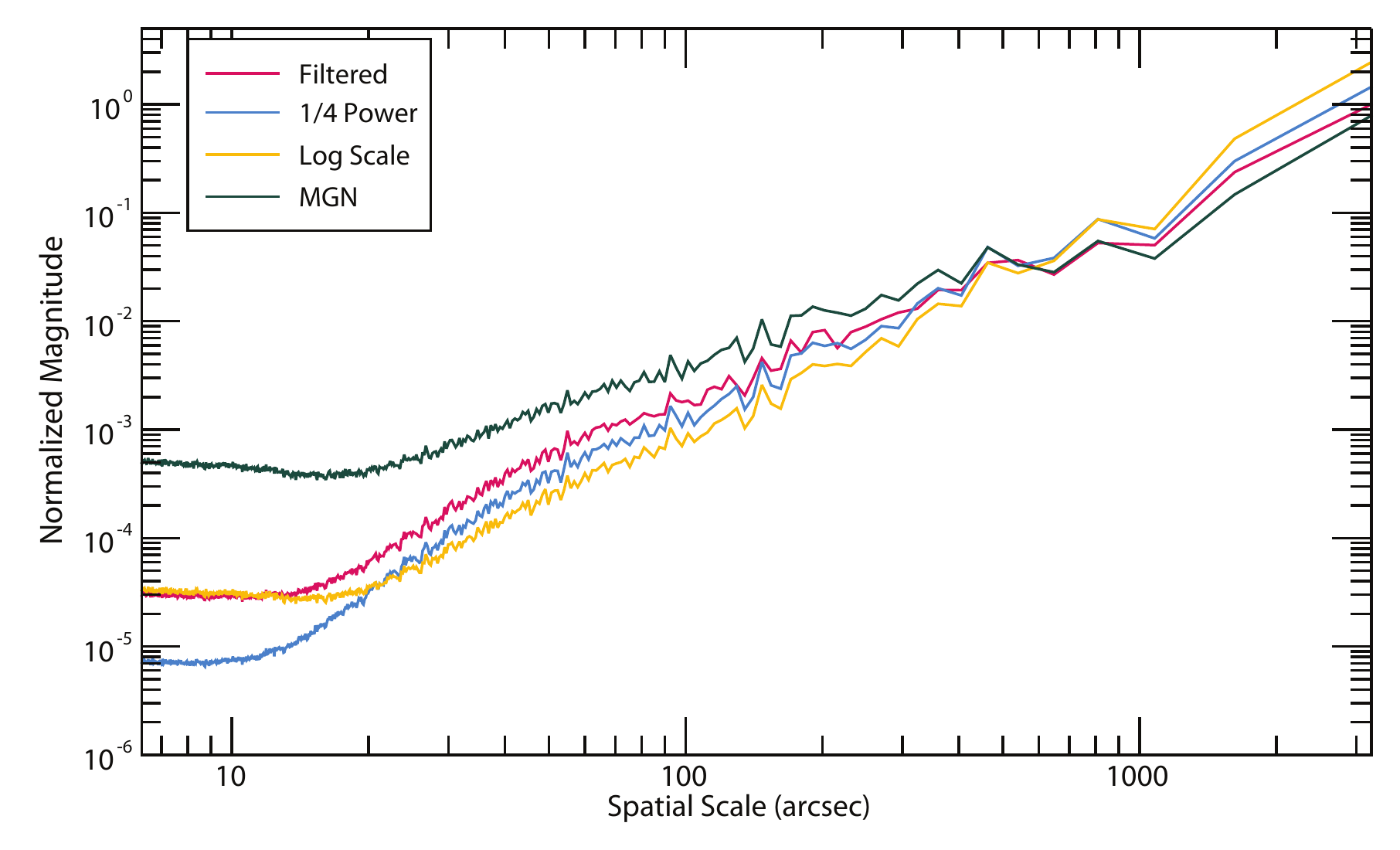}
\caption{Azimuthally averaged power spectra of images processed with the SWAP filter compared to log-scaling, \sfrac{1}{4}-Power image scaling, and MGN processing.}\label{fig:power_spec}
\end{figure}

To more quantitatively highlight differences in the SWAP filter's performance compared to other image scaling and filtering techniques, we compute the azimuthally averaged power spectrum for the images processed with the SWAP filter (see Figure~\ref{fig:filter_example}) and compare the resulting spectrum to those of the log and \sfrac{1}{4}-power scaling (see Figure~\ref{fig:scaling}), as well as the MGN filter (see Figure~\ref{fig:mgn_nrgf}); the results are shown in Figure~\ref{fig:power_spec}. (Such power spectrum analyses have been used before to show similarities and differences between solar images where the spatial frequencies are known to differ, such as by \citet{Rachmeler2019}, Figure~10.)

The MGN enhances structures on relatively small scales, and thus has the most power in this range, but it fully suppresses the largest-scale structures, and has the least power for these scales. The \sfrac{1}{4}-power, log-scaled, and SWAP-filtered images are all are similarly behaved, with the SWAP filter emphasizing small and mid-range scales, and thus having the most power in this range. The log-scaled image suppresses the mid-range frequencies, but equalizes brightness across the whole image, and thus has the most power at large scales. Surprisingly, the log-scaled and SWAP-filtered images are very similarly behaved at high frequencies. In fact, this appears to be because they enhance off-disk noise -- a very high-frequency feature -- compared to the \sfrac{1}{4}-power filter, which attenuates the off-disk brightness much more than these other methods.

Overall, these power spectra show more quantitatively what we have, up until now, mainly illustrated qualitatively: the SWAP filter results in images that more closely resemble scaled, but otherwise unprocessed, solar images, while also improving the image contrast over these simpler methods for features across a wide range of scales.

\section{Conclusions}
\label{sec:conclusions}

Here we have described the SWAP Filter: a newly available azimuthally varying radial normalizing filter, specifically designed for wide-field EUV images of the solar corona. Although simple, the filter is highly effective for producing displayable EUV images that preserve both large-scale EUV features and small-scale variations within the global coronal structure. The filter has been demonstrated on a large variety of EUV image data, and is now publicly available via the PROBA2/SWAP \textsf{SolarSoft IDL} software package. Performance tests using an 2.40 GHz Intel Core i9-9980HK demonstrate that filter generation with default parameters typically takes less than 0.3~s for a $1024\times1024$ SWAP image, so the technique is fast and computationally light compared to other similar techniques, making this a useful approach when computational resources are limited or for large datasets.

Image processing tools such as this one are especially important for studies of the middle corona, as observations that connect this region to the solar disk inherently contain very large brightness gradients that must be reduced in order to effectively display the entire corona. Numerous approaches are available, each of which offers various advantages for specific applications. Our filter is fast to apply and it yields images that do not appear extensively processed, and is well suited for the display of data that emphasizes global-scale features, and for generating outreach images and movies that resemble the appearance of inner coronal images that have been available for decades, and thus feel familiar to viewers.

%%%%%%%%%%%%%%%%%%%%%%%%%%%%%%%%%%%%%%%%%%%%%%%%%%%%%%%%%%%%%%%%%%%%%%%%%%%
\begin{acknowledgments}
  SWAP is a project of the Centre Spatial de Liège and the Royal Observatory of Belgium funded by the Belgian Federal Science Policy Office (BELSPO). SWAP observations appearing in this article are available via the Virtual Solar Observatory or the PROBA2 website via \url{https://proba2.sidc.be/}. SWAP software tools, including the SWAP Filter, are available via \textsf{SolarSoft IDL}. SUVI observations shown in this article are available via NOAA NCEI's Space Weather data website, accessible via \url{https://doi.org/10.25921/D60Q-G238}.

 D.B.~Seaton and M.J.~West acknowledge support from NASA’s Heliophysics Guest Investigator program, grant 80NSSC22K0523.
\end{acknowledgments}

%\noindent To change a title use an optional parameter:\par
%\verb+\begin{acks}[Acknowledgements]...\end{acks}+

%\acknowledgment US spelling: \verb+\acknowledgment+
%\acknowledgement British  spelling: \verb+\acknowledgement+

%%%%%%%%%%%%%%%%%%%%%%%%%%%%%%%%%%%%%%%%%%%%%%%%%%%%%%%%%%%%%%%%%%%%%%%%%%%
%\appendix   

% If there wer an appendix, there would be text here

%%% BIBLIOGRAPHY %%%%%%%%%%%%%%%%%%%%%%%%%%%%%%%%%%%%%%%%%%%%%%%%%%%%%%%%%%%

     % format of references provided by the journal (.bst)
\bibliographystyle{spr-mp-sola}
     % name your Bibtex file containing your references (.bib)
\bibliography{refs.bib}

\end{article} 

\end{document}